\documentstyle[amsfonts,amsmath,graphicx,verbatim,slashed,12pt]{article}

\newcommand {\slsh} [1] {\not{\hbox{\kern-2pt${#1}$}}}

\def\drawbox#1#2{\hrule height#2pt
         \hbox{\vrule width#2pt height#1pt \kern#1pt
               \vrule width#2pt}
               \hrule height#2pt}

\def\Asym#1#2{\vcenter{\vbox{\drawbox{#1}{#2}
               \kern-#2pt       
               \drawbox{#1}{#2}}}}

\newcommand {\beq} {\begin{equation}}
\newcommand {\eeq} {\end{equation}}
  \newcommand {\ber}{\begin{eqnarray*}}
  \newcommand {\eer} {\end{eqnarray*}}
\newcommand {\bea}{\begin{eqnarray}}
  \newcommand {\eea} {\end{eqnarray}}

\newcommand{\Dslash}{\,{\raise.15ex\hbox{/}\mkern-12mu D}}

\begin{document}


\begin{titlepage}

\begin{center}
\vspace{1in}
\large{\bf Level-Rank Duality in Chern-Simons Theory from}\\
\large{\bf a Non-Supersymmetric Brane Configuration }\\
\vspace{0.4in}
\large{Adi Armoni and Edwin Ireson}\\
\small{\texttt{a.armoni@swan.ac.uk , pyireson@swansea.ac.uk}}\\
\vspace{0.2in}
\emph{Department of Physics, Swansea University}\\ 
\emph{Singleton Park, Swansea, SA2 8PP, UK}\\
\vspace{0.3in}
\end{center}

\abstract{We derive level-rank duality in pure Chern-Simons gauge theories from
a non-supersymmetric Seiberg duality by using a non-supersymmetric brane configuration in type IIB string theory. The brane configuration consists of fivebranes, $N$ D3 antibranes and an O3 plane. By swapping the fivebranes we derive a 3d non-supersymmetric Seiberg duality. After level shifts from loop effects, this identifies the IR of $Sp(2N)_{2k-2N+2}$ and $Sp(2k-2N+2)_{-2N}$ pure Chern-Simons theories, which is a level-rank pair. We also derive level-rank duality in a Chern-Simons theory based on a unitary group.}

\end{titlepage}

\section{Introduction}

Level-rank duality between two Chern-Simons theories has been a well known relation for many years \cite{Naculich:1990hg}. It has recently been found that it is closely related to 3d Seiberg duality \cite{Seiberg:1994pq,Aharony:2008gk,Giveon:2008zn,Armoni:2009vv,Kapustin:2011gh}. The 3d Seiberg duality had been motivated by a realization of the gauge theory on a type IIB brane configuration and the fivebrane swapping procedure \cite{Aharony:2008gk,Giveon:2008zn,Elitzur:1997fh}.

So far the study of 3d Seiberg duality for Yang-Mills Chern-Simons theories was restricted to supersymmetric theories with various degree of supersymmetry (the case with minimum supersymmetry was studied in \cite{Armoni:2009vv}). Breaking supersymmetry is hard and typically leads to instabilities. Moreover, in the absence of holomorphicity it is difficult to prove or even motivate the duality. The case of level-rank duality in pure Chern-Simons theory is exceptional in that respect, since the duality is known to hold in the absence of supersymmetry. 

In this paper we consider non-supersymmetric brane configurations that consist of anti-branes and an orientifold plane, also known as Sugimoto model \cite{Sugimoto:1999tx}. While this class of brane configurations breaks supersymmetry explicitly, one may hope that it preserves properties such as S-duality\cite{Sugimoto:2012rt,Hook:2013vza} and Seiberg duality \cite{Armoni:2013ika,Armoni:2013kpa}. In particular, in this paper we consider a certain brane configuration that leads to a non-supersymmetric Seiberg duality. We show that in the IR this duality becomes the celebrated level-rank duality between two Chern-Simons theories based on $Sp$ gauge group
 \beq
  Sp(2N)_{2k-2N+2}\,\,\, \sim \,\,\, Sp(2k-2N+2)_{-2N} \, . \label{duality}
\eeq

In addition, we also comment on non-supersymmetric Seiberg duality and level-rank duality between theories based on a unitary gauge group. In this case we use a type 0B brane configuration to support our claim.

The organization of the paper is as follows: in section 2 we present the brane setup and propose a non-supersymmetric Seiberg duality. In section 3 we study the dynamics of the gauge theory. In section 4 we consider the case of a unitary group. Finally, in section 5 we discuss our results.

\section{Brane configuration and duality}

We study brane configuration similar to those that give rise to 3d ${\cal N}=2$ Yang-Mills Chern-Simons theories \cite{Giveon:2008zn}. The brane configuration consists of 

\begin{itemize}
\item an NS5 brane along the $012345$ directions,
\item an $O3$ plane along the $0126$ directions and $N$ {\it anti} D3 branes (and their mirrors) along the $012|6|$ directions,
\item a tilted $(1,2k)$ fivebrane along $012(37)89$ directions. The latter fivebrane is a bound state of an NS5 brane and $2k$ {\it anti} D5 branes, it is tilted in the $(37)$ plane by an angle $\theta$ such that $\tan\left( \theta\right)=-2g_sk$. With this choice of angle, we would preserve ${\cal N}=2$ supersymmetry, had we not had an orientifold. The minus sign reflects the presence of anti-branes in the construction.
\end{itemize}  The orientifold plane is of $O3^+$ type in between the fivebranes. When it crosses the fivebranes it becomes $O3^-$. The brane configuration is depicted in figure \eqref{electric} below. We refer to it as the ``electric'' theory.

Despite of the fact that the brane configuration breaks supersymmetry explicitly, it is stable. As we shall see, there is an attractive potential between the $O3$ plane and the anti $D3$ branes that ensures that the branes will not move away\footnote{An instability develops in the case where the orientifold type between the fivebranes is $O3^-$. Hence, a statement similar to ours, for an $SO$ gauge group, is not obtainable immediately by our method.}. 

\begin{figure}[!ht]
\centerline{\includegraphics[width=6cm]{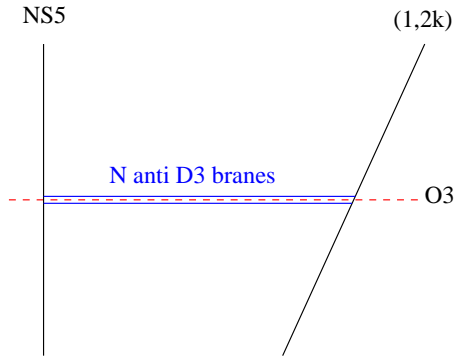}}
\caption{\footnotesize The electric theory. It is a non-supersymmetric $Sp(2N)$ Yang-Mills theory with a level $2k$ Chern-Simons term. In this plot $N=1$.}
\label{electric}
\end{figure}
The gauge theory on the electric brane configuration is a $Sp(2N)$ gauge theory, with two Majorana fermions $\lambda$ and $\psi$  that transform in the two-index antisymmetric representation and an additional real scalar $C$ that transforms in the two-index symmetric representation of the gauge group. Changing the fermionic representation breaks supersymmetry explicitly, this is a consequence of taking anti-branes with an orientifold, as in \cite{Sugimoto:2012rt,Hook:2013vza,Armoni:2013ika}.

 The Lagrangian of the theory is given by
\bea
& &
{\cal L}= \label{lagrangian} \\
& & {1\over g^2} \text{Tr}\{ -{1\over 2} F_{\mu \nu}^2 + (D_\mu C)^2 +D^2+ i \bar \lambda \slashed{D} \lambda + i \bar \psi \slashed{D} \psi  + i\left[ \bar{ \psi},\lambda\right] C  -i \left[\bar{ \lambda},\psi\right] C \}  \nonumber \\
& &+ 2k \text{Tr}\{ \epsilon ^{\mu \nu \rho} (A_\mu \partial _\nu A_\rho -{2\over 3} iA_\mu A_\nu A_\rho ) - \bar \lambda \lambda + \bar \psi \psi + 2CD  \} \, , \nonumber
\eea
with $g$ the Yang-Mills coupling and $2k$ the Chern-Simons level.

Since we are mixing representations of the gauge group, it is easier to work with fields in the double-index notation. Instead of expanding the fields in the relevant bases of the Lie algebra, we keep them to be colour matrices, with (anti-)symmetrised indices. Traces are then taken over these colour indices to produce terms in the Lagrangian.

 According to the proposal of Elitzur-Giveon-Kutasov \cite{Elitzur:1997fh}, by swapping the fivebranes we arrive at the magnetic Seiberg dual of the electric theory. When we swap fivebranes in the presence of an $O3$ brane a single $D3$ antibrane (and its image) is created \cite{Aharony:2008gk}. This phenomenon is often attributed to the conservation of a particular quantity, calculated from various brane charges, called Linking Number \cite{Hanany:1996ie}. Together with the $k$ antibranes and the $N$ branes the magnetic gauge group becomes $Sp(2k-2N+2)$. The magnetic theory is depicted in \eqref{magnetic} below.
\begin{figure}[!ht]
\centerline{\includegraphics[width=6cm]{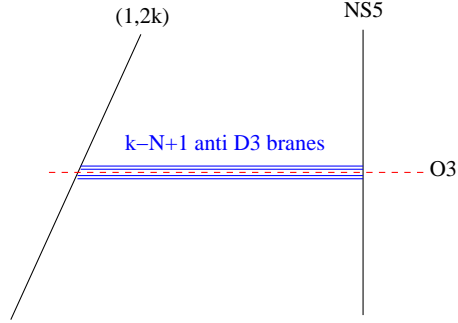}}
\caption{\footnotesize The magnetic theory. It is a non-supersymmetric $Sp(2k-2N+2)$ Yang-Mills theory with a level $-2k$ Chern-Simons term. In this plot $k-N+1=2$.}
\label{magnetic}
\end{figure}

Thus we propose that two non-supersymmetric gauge theories form a Seiberg dual pair, namely that the two theories become equivalent in the far IR, based on string theory dynamics. These two theories are Yang-Mills Chern-Simons, both admit a level $|2k|$ at the classical level. The electric theory is based on the gauge group $Sp(2N)$ and the magnetic on $Sp(2k-2N+2)$. In the next section we will argue that in the far IR the electric and magnetic theories flow to a pair of pure Chern-Simons theories that are equivalent to each other by virtue of level-rank duality.

\section{Gauge dynamics}

Let us consider the gauge dynamics of the electric and magnetic theories.

We start with the electric theory. Since the theory is non-supersymmetric the scalar is expected to acquire a Coleman-Weinberg potential. Let us consider the one-loop contribution to the scalar mass, as depicted in figure \eqref{mass} below.

\begin{figure}[!ht]
\centerline{\includegraphics[width=8cm]{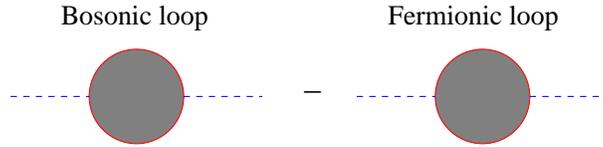}}
\caption{\footnotesize Perturbative contributions to the scalar mass.}
\label{mass}
\end{figure}

The contributions from bosonic one-loop diagrams are
\beq
 +g^2 (2N+2) \int {d^3 p\over (2\pi)^3} {1\over p^2} =  c g^2 (2N+2) \Lambda \, ,
\eeq
with $g$ the gauge coupling, $\Lambda$ the UV cut-off and c some numerical constant. The contributions from fermionic one-loop diagrams are 
\beq
 -g^2 (2N-2) \int {d^3 p\over (2\pi)^3} {1\over p^2} = -c g^2 (2N-2) \Lambda \, .
\eeq
The planar diagram contributions (dependent on N) cancel, but the difference between the contributions of the non-planar diagrams, due to the difference in representations between bosonic and fermionic degrees of freedom, do not cancel. The generated mass for the scalar is therefore
\beq
M^2 = cg^2 \left \{ (2N+2) - (2N-2) \right \} \Lambda = 4cg^2 \Lambda \, .
\eeq 
Thus, a positive mass$^2$ proportional to the UV cut-off is generated for the scalar and therefore it decouples from the dynamics of the low-energy field theory. 

In the brane side, a potential of the form $M^2 C ^2$ is interpreted as an attractive potential between the orientifold $O3^+$ plane and the D3 antibranes. It guarantees the stability of the brane configuration\footnote{In the case of $O3^-$, the gauge theory is $SO(2N)$ and the scalar develops a tachyonic mass. The brane configuration, hence, becomes unstable.}.

The field theory well below the cut-off scale $\Lambda$ is a Chern-Simons theory coupled to two Majorana antisymmetric fermions. Such theories, when non-supersymmeric, typically undergo a finite shift in level, due to loop corrections to the gluon propagator. The analysis of \cite{Kao:1995gf} shows notably that of the seven Feynman diagrams involved in this process, only three yield a non-zero contribution, one gluonic and two fermionic. The total shift can then be calculated to be (in our case)
\beq
2k \rightarrow 2k-(2N-2) \, .
\eeq
In particular, the scalars do not contribute to this shift. We are therefore justified in our procedure: it is natural to decouple the scalars first, as their mass is a stringy correction, and materialises the stability of our brane construction. This does not affect subsequent computations.

In addition to this, both the gauge bosons and the gluon acquire a Chern-Simons mass $M \sim g^2 k$, such that the IR theory becomes the topological pure $Sp(2N)$ Chern-Simons theory with level $2k-2N+2$.\footnote{This level will be further shifted due to a gluonic loop by $2N+2$, namely $2k-2N+2 \rightarrow 2k+4$.}

The same analysis for the magnetic theory can be made. It has a negative level, but level shifts depend on this sign explicitly: the IR theory in this case is the pure $Sp\left(2k-2N+2\right) $ theory, with shifted level
\beq
 -2k \rightarrow -(2k - ((2k-2N+2)-2))=-2N \, ,
\eeq
 as required.

The electric and magnetic theories are known to be dual to each other \eqref{duality}. We, thus, recover the famous level-rank duality from a non-supersymmetric setup in string theory.

\section{Unitary gauge group}

Similarly to the previous analysis, we may start with type 0B string theory. The brane configurations are identical to those that are depicted in plots \eqref{electric} and \eqref{magnetic}. In type 0B string theory there are three type of orientifolds. We consider a setup with the orientifold $O3^+$ plane of Sagnotti\cite{Sagnotti:1995ga,Sagnotti:1996qj}, such that the closed string tachyon is projected out from the worldvolume of the D3 antibranes. The orientifold also projects out half of the doubled set of RR fields of type 0B. The RR forms, and hence the D3 branes, are as in type IIB. The NSR sector is absent and there are no closed string fermions. The resulting string theory is non-supersymmetric and shares a lot of similarities with the Sugimoto model.   

 The electric theory on the branes is a level $2k$ $U(2N)$ Yang-Mills Chern-Simons theory with a scalar in the adjoint representation and two Majorana fermions that transform in the two-index antisymmetric representation. It is described by the Lagrangian \eqref{lagrangian}.

By swapping the fivebranes we arrive at the magnetic gauge theory. The swapping procedure in type 0 string theory can be justified by using worldsheet argument. The magnetic gauge dual is a level $-2k$ $U(2k-2N+2)$ Yang-Mills Chern-Simons theory, described by the same Lagrangian \eqref{lagrangian}, based on a unitary group.

As in the previous case with $Sp$ gauge group, the scalar acquires a positive mass$^2$ and decouples from the low energy dynamics. The IR theories, after the shift due to the fermions is taken into account, become the pure bosonic Chern-Simons theories. The IR electric theory is $U(2N)_{2k-2N+2}$. The IR magnetic theory is $U(2k-2N+2)_{-2N}$. The two theories are well known to be equivalent to each other
\beq
 U(2N)_{2k-2N+2}\,\,\, \sim \,\,\, U(2k-2N+2)_{-2N} \, , \label{ulr}
\eeq
due to level-rank duality \cite{Naculich:1990hg}. Thus, despite of the absence of supersymmetry in the underlying string theory, we recovered the celebrated result \eqref{ulr}.

\section{Summary}

In this paper we considered a non-supersymmetric brane configuration that leads to a non-supersymmetric Seiberg duality. Usually, non-supersymmetric configurations in string theory cannot be trusted, since they are potentially unstable.

The class of configurations that we considered is special. It contains no open or closed string tachyons, both in the electric and in the magnetic side of the duality. Moreover, quantum corrections shift the square mass of the field theory scalar towards positive values, hence guarantee the stability of the configuration.

One may still doubt whether the fivebrane swapping argument can be trusted in the absence of supersymmetry. Indeed, even in the supersymmetric case, it is not clear {\it why} swapping the fivebranes should lead to a Seiberg duality. In fact, the main motivation for this study is to check whether Seiberg duality holds for this particular class of non-supersymmetric gauge theories.

The result is very encouraging, because the duality is genuinely non-supersymmetric. The IR field theory ``knows'' that the far UV is non-supersymmetric. The shifted level of the Chern-Simons term depends on the representation of the fermions. It is crucial, for the level-rank duality to hold, that the fermions transform in the antisymmetric representation. It is also crucial that the dual gauge group is $Sp(2k-2N+2)$ -- and that happens only when there is an orientifold and {\it antibranes}. Thus, even without supersymmetry all pieces fall into the right place!

In this paper we focused on $Sp$ and $U$ gauge groups with antisymmetric fermions. The cases of $SO$ and $U$ gauge groups with symmetric fermions, are less understood. In these cases the brane configuration is unstable and it is difficult to find the true vacuum of the theory. The most plausible scenario is that the scalar acquires a vev, leading to a purely bosonic Chern-Simons theory in the true vacuum.

Satisfied by our results, we hope to consider more complex theories in future works.

\vskip 0.5cm
{\it \bf Acknowledgments.} We wish to thank professor Prem Kumar for a useful discussion. 


\end{document}